\documentclass[letterpaper]{article}
\usepackage{amsfonts, latexsym, amssymb, amsthm, algorithmic,comment,amsmath}
\usepackage{times} 
\usepackage{helvet} 
\usepackage{courier}\title{Title}

\newcommand{\swap}{\circ}

\renewcommand{\setminus}{\mathbf{-}}
\newcommand{\deficit}{\mathit{Deficit}}

\renewcommand{\P}{\mathcal{P}}
\newcommand{\NP}{\mathcal{NP}}

\newcommand{\w}[2]{#1(#2)}

\newcommand{\dfr}[3]{D(#2,#1,#3)}
\newcommand{\tab}{~~~~~}

\newcommand{\bigO}{O}

\newcommand{\bigOmega}{\Omega}
\DeclareMathOperator*{\argmin}{argmin}
\DeclareMathOperator*{\argmax}{argmax}

\newtheorem{definition}{Definition}
\newtheorem{theorem}[definition]{Theorem}

\newtheorem{proposition}[definition]{Proposition}

\begin{document}
\title{Efficient, deterministic voting rules that approximate Dodgson and 
Young scores}
\author{Jason A. Covey and Christopher M. Homan}
\maketitle

\begin{abstract}
\begin{quote}We provide deterministic, polynomial-time computable voting rules 
that 
approximate Dodgson's and (the ``minimization version'' of) Young's
scoring rules to within a logarithmic 
factor. Our approximation of Dodgson's rule is tight up to a constant
factor, as Dodgson's rule is $\NP$-hard to approximate to within some 
logarithmic factor. The ``maximization version'' of Young's rule is known to be 
$\NP$-hard to approximate by any constant factor.
Both approximations are simple, and natural
as rules in their own right: Given a candidate we wish to score, we can regard 
either its Dodgson or Young score as the edit distance between a given
set of voter preferences and one in which the candidate to be scored is the 
Condorcet winner. (The difference between the two scoring rules is the type of
edits allowed.) We regard the marginal cost of a sequence of edits to
be the number of edits divided by the number of 
reductions (in the candidate's deficit against any of its opponents 
in the pairwise race against that opponent) that the edits yield. Over 
a series of rounds, our scoring rules greedily choose a sequence of edits that 
modify exactly one voter's preferences and whose marginal cost is no greater 
than any other such  single-vote-modifying sequence.
\end{quote}
\end{abstract}
\section{Introduction}
A voting rule takes a collection of voter preferences (over
some fixed set of candidates, or alternatives) and aggegrates them
into a single ranking, ideally in a way that is as ``fair'' as possible
to the voters. 
Arrow's famous impossibility theorem~\cite{arr:j:difficulty} states that no such rule
over three or more candidates meets all
reasonable fairness criteria. So when considering such
rules it may be important to know which criteria they do and do not meet.

One such criterion, which actually predates those studied explicitly by 
Arrow, is credited to the Marquis de 
Condorcet~\cite{con:b:condorcet-paradox}.\footnote{In fact, centuries 
earlier Lull considered essentially 
the same criterion~\cite{hae-puk:j:electoral-writings-ramon-llull}.} A 
candidate that, against any opposing candidate, is
preferred by a majority of voters is called the Condorcet winner. Note
that different majorities may prefer the Condorcet winner to
different opponents. Note also that such a winner may not
exist, but if it does then it is unique.
 The Condorcet criterion states that, whenever a Condorcet
winner does exist, it must be declared the winner. See~\cite{you:j:extend}
for a nice discussion of its virtues.

Unfortunately, many widely used rules, such as plurality, instant-runoff,
and Borda count do not have this very natural property. Many that do bring 
with them undesirable features, for instance
Copeland elections~\cite{cop:u:reason} tend to frequently result in ties. 
Others, 
such as those due to Dodgson~\cite{dod:unpubMAYBE:dodgson-voting-system}, 
Kemeny~\cite{kem:j:math}, and Young\footnote{Kemeny's 
voting rule is sometimes called the Kemeny-Young rule, as Young studied it 
and made some important breakthroughs~\cite{you-lev:j:consist,you:j:condorcet},
e.g., he showed that it satisfies the Condorcet criterion.
The Young-only rule to which we refer is distinct from the Kemeny-Young rule, which to
avoid confusion we will call simply ``Kemeny.''}~\cite{you:j:extend} are $\NP$-hard to 
compute~\cite{bar-tov-tri:j:who-won,jor-spa-vog:j:win-young}. In fact, they are complete with respect 
to parallel access to 
$\NP$~\cite{hem-hem-rot:j:dodgson,hem-spa-vog:j:kemeny,jor-spa-vog:j:win-young}, which means that even if the problem of determining 
the winner according to one these rules is ``merely'' in $\NP$, the polynomial
hierarchy would collapse (to $\NP$).

We can view Dodgson's and Young's rules as variations on a theme:
Given a list, or profile, of the voters' preferences 
(here as is standard in the theory of voting we take each voter's preferences to 
be a total ranking over all the 
candidates) and a candidate we wish to score, either rule
takes as the candidate's score the edit distance~\cite{cor-lei-riv-ste:b:algorithms-second-edition} between the given
preference profile and one that makes the candidate a Condorcet winner.
In other words, it is the number of edits (exactly what an edit is depends on
the particular scoring rule) needed to reduce to zero the vote deficit between
the given candidate and each of its rivals.
Candidates are then ranked in increasing order by their scores.
 In Young's rule, an edit simply
deletes one voter from the list. For Dodgson, an edit 
takes one voter's ranking and replaces it with one
just like it, except that in the new one the positions of
one pair of candidates ranked adjacently in the original list are swapped.
Clearly, both rules satisfy the Condorcet criterion, as any Condorcet
winner has a score of zero.

A simple example illustrates how scoring works. Let $a$, $b$, $c$, $d$, and
$e$ be five candidates and let
\begin{eqnarray*}
  &a>_1b >_1 c >_1 d >_1 e&\\
  &a>_2b >_2 c >_2 d >_2 e&\\
  &d>_3a >_3 e >_3 c >_3 b&\\
  &d>_4a >_4 e >_4 c >_4 b&\\
  &c>_5e >_5 b >_5 d >_5 a&\\
\end{eqnarray*}
be a preference profile having five voters. In this example, no candidate
is a Condorcet winner. Note that $c$ is preferred over $b$, $d$, and $e$ by
majorities of voters and is losing to $a$ by four votes. To make $c$ the 
Condorcet winner, we
could swap $c$ with $b$ and then with $a$ in voter one and two's rankings. 
It turns
out there is no shorter sequence of swaps that makes $c$ the Condorcet winner,
so the Dodgson score of $c$ is four.
Note that in this case, the swaps between $c$ and $b$ do not actually reduce
$c$'s vote deficit, since $c$ is already beating $b$. 

Candidate $d$ is losing to $c$ and $b$ by one vote each. To make $d$
the Condorcet winner, we could remove voters one and two. Thus $d$'s Young 
score is two. In this case, both removals yield two deficit reductions, but
in general the number of deficit reductions that each removal 
yields will vary.
 
As Procaccia et al.~observe~\cite{pro-fel-ros:t:approx-innapprox-d-y},
McCabe-Dansted effectively proves that it is hard to
$\bigOmega(\log m)$-approximate Dodgson elections, where $m$ is the 
number of candidates~\cite{mcc:t:approx-comp}. In the same paper,
Procaccia et al.~show that it is hard to approximate the ``maximization
version'' of Young's score---i.e., where the Young 
score is taken to be the largest subset
of voters that makes a given candidate the Condorcet winner\footnote{This
is the actual definition due to Young~\cite{you:j:extend}. Our formulation 
in terms of deletions is used elsewhere
(see, e.g.,~\cite{jor-spa-vog:j:win-young,bet-guo-nie:u:param,fishburn:j:condorcet}), and is in 
many respects
equivalent to the original definition (though certainly not with
respect to optimization and approximation results, at least not
directly). Moreover, the 
deletion-based version we use allows us to more naturally build
Dodgson's and Young's rules into a single framework.}---by any constant
factor~\cite{pro-fel-ros:t:approx-innapprox-d-y}.

In this paper we present a framework for efficient, edit-based scoring 
rules. From this framework, we obtain $\bigO(\log m)$ 
approximations of the scoring rules due to Dodgson and Young. The
basic idea is very simple: Given a profile of voter preferences
and a candidate we wish to score, let the marginal cost of a
sequence of edits be the number of edits divided by the number
of times that, as the edits are applied, the vote deficit
against the candidate we wish to score is reduced. Now,
proceed over a series of rounds to edit the 
profile until the chosen candidate becomes the Condorcet winner. In each
round, greedily choose a voter and a sequence of edits on that voter's
preferences that, over all such voters and sequences, has the
minimum marginal cost. 

It turns out that, when we restrict the edits the algorithm 
makes to those allowed by Dodgson's (respectively,
Young's) scoring rule, the result is a polynomial-time
$\bigO(\log m)$-approximation,
where $m$ is the number of candidates.
Thus, in the case of Dodgson elections, the approximation
is tight up to a constant factor. 
 
Why care about approximations to voting rules in the first place? One reason
is that they are themselves voting rules, ones 
that in some way relate to the rules they approximate. We feel that 
our framework supplies approximations that are simple and natural 
enough to function as voting rules in their own 
right. 

For instance, suppose a group of voters agrees to only 
accept a Condorcet winner. If their stated individual preferences
fail to yield one, then the election controller holds an 
auction, to entice some of the voters to change their minds.

Taking one candidate at a time, the 
controller offers to pay each voter for each reduction in the
candidate's vote deficit it can deliver by changing its stated preferences.
The cost to the voter is the number of edits it needs to make.
If the price offered is less than the cost to the voter, the voter will not
accept. If not enough voters accept, 
the controller increases the amount offered and the process repeats until
the candidate becomes the Condorcet winner.
The score of the candidate is then the total amount of money offered to the
voters and the candidate having the lowest score is the winner.
(No payoffs occur until after all candidates are 
scored, and only those deals made during the winning candidate's scoring
round are actually honored, so in effect the voters ``choose'' a
Condorcet winner.)

The auction thus encourages
voters to reveal the true value of their edits, as those who
are willing to take the least amount of money per deficit reduction delivered
are rewarded first, while those holding out for more may get nothing.
Assuming that all voters uniformly value their
edits at some common unit price, the score the auction provides
(and the order in which it selects the swaps to make) coincides with our
rules.

\subsection*{Related work}
The study of the approximibility of voting rules is rather new.
Ailon et al.~\cite{ail-char-new:c:agg},
Coppersmith et al.~\cite{cop-fle-rud:c:order}, and
Kenyon-Mathieu and Schudy~\cite{ken-sch:c:how}
study approximation algorithms on Kemeny elections.

As noted above, McCabe-Dansted~\cite{mcc:t:approx-comp} (respectively, 
Procaccia~\cite{pro-fel-ros:t:approx-innapprox-d-y})
provides lower (respectively, upper) bounds on approximating Dodgson 
(respectively, Young) 
scores. Additionally, Procaccia et al.
provide a polynomial-time, \emph{randomized} algorithm that with probability
at least 1/2 $\bigO(\log m)$-approximates the Dodgson 
score~\cite{pro-fel-ros:t:approx-innapprox-d-y}.
They use a linear program whose optimal solution may assign
fractional values to counts of the swaps made. They then
use randomness to help assign integer values to the swap counts, in a
way that yields
a feasible, integer-valued solution. Our results improve 
on this approach in that our algorithm is completely deterministic and,
we feel, more straightfoward and natural. Additionally, we provide
a polynomial-time approximation of Young scores.

Several researchers provide algorithms that run in polynomial time
on key subsets of the problem domain. Bartholdi et 
al., in the same seminal paper that established
$\NP$-hardness results for Dodgson and Kemeny 
elections~\cite{bar-tov-tri:j:who-won},
show that Dodgson elections can be scored in polynomial time when
either the number of candidates or the number of voters is fixed.
Our algorithm runs in polynomial time on all inputs, however it
is does not guarantee to provide a correct answer. Rather, it
guarantees upper bounds on the degree of error.

Homan and Hemaspaandra~\cite{hom-hem:j:guarantees} and 
McCabe-Dansted et al.~\cite{mcc-pri-sli:j:approximability-of-dodgson} 
use a common insight to provide polynomial-time, deterministic 
heuristics that, in cases where
the voters greatly outnumber the candidates, compute with
high probability the exact Dodgson score on a candidate and preference profile 
chosen uniformly at random from all profiles of some fixed size. Our 
Dodgson-score-approximating algorithm is a generalization of sorts of
their approach. Though we do not analyze the probability of exactness our
algorithm has, we note here that whenever the Homan and Hemaspaandra approach 
correctly computes the Dodgson score, so does ours. However, when their 
algorithm is not exact, it returns a score that is less than the true edit 
distance. Our algorithm never
returns a score that is less than the edit distance. Moreover, our algorithm
 always builds as a side effect an actual sequence of edits leading to a Condorcet
 winner.

Finally, Rothe et al.~(in the same paper where they establish optimal
bounds on the complexity of Young elections) give a polynomial-time
algorithm for computing the ``homogeneous'' versions 
(see~\cite{fishburn:j:condorcet}) of Dodgson's and Young's voting 
rules~\cite{jor-spa-vog:j:win-young}. (A voting rule is homogenous if cloning
each voter's preferences some fixed number of times does not affect the 
score). They do not discuss the degree to which these scores
approximate Dodgson and Young rules.

\section{Definitions}
\subsection{Elections}
Let $V=\{1,\ldots,n\}$ be a set of voters and $C$ be
a set $\{1,\ldots,m\}$ of candidates. A ranking of the candidates is a total
ordering over $C$, i.e., $\langle c_m > c_{m-1}> \cdots > c_1\rangle$, where 
$\{c_1,\ldots,c_m\} = C$.
We denote the set of all such rankings $\mathcal{L}(C)$.
The voters' preference profile is an $n$-tuple in $\mathcal{L}(C)^n$.
 For a given preference profile $\langle >_1,\ldots,>_n\rangle \in \mathcal{L}(C)^n$, $i \in V$, and $c\in C$, let $\w{c}{>_i}$
denote $||\{d \in C~|~c >_i d\}||.$

For every pair of distinct candidates $c,d \in C$ and every preference profile 
$P= \langle >_1,\ldots, >_n \rangle$, $c$'s vote deficit in $P$ with $d$ is 
$\deficit(P,c,d) = \min\{0, ||\{i \in V~|~d >_i c\}|| - ||\{i \in V~|~c >_i d\}||\}.$
The \emph{total deficit} of $c$ is 
$$\deficit(P,c) = \sum_{d \in C \setminus\{c\}} \deficit(P,c,d).$$ 
Thus $c$ is a \emph{Condorcet winner} if and only if $\deficit(P,c) = 0$.
$\deficit(P,c)$ is sometimes known as the Tideman score~\cite{tid:j:clones-dodgson}, which forms the basis of the Tideman (a.k.a., ranked pairs) voting
rule.
\subsection{Edit-based scoring rules}
The building blocks of this paper are edits and deficit reductions. It
will be useful to view them as objects we can label. We now show
how to do this.

An edit is a mapping $e:\bigcup_{i=0}^\infty \mathcal{L}(C)^i 
\rightarrow\bigcup_{i=0}^\infty  \mathcal{L}(C)^i$. Let $P\swap e$ denote the 
application of $e$ to some preference profile $P$.
A sequence of edits $\langle e_1,\ldots, e_p\rangle$ is called a Condorcet
sequence if $\deficit(P\swap e_1\swap \cdots \swap e_p,c) = 0$

A swap is an edit, designated 
by an ordered pair $(i,j)\in\mathbb{N}^2$, that takes a preference profile 
$P=\langle >_1,\ldots,>_n\rangle$ and outputs 
$\langle >_1',\ldots,>_n'\rangle$, which is
just like $P$ except that, if $1\leq i \leq n$ and $0< j < m$,
then for $c,d\in C$ satisfying $\w{d}{>_i} = j = \w{c}{>_i} + 1$ it holds that
 $\w{c}{>'_i} = j = \w{d}{>'_i}+1$, i.e., $c,d$ are adjacent in both rankings,
$d >_i c$, and $c >'_id$.
Candidates $c$ and $d$ are said to be involved in the swap.

A deletion is an edit, designated by some $i \in \mathbb{N}$, that takes a preference profile 
$P=\langle >_1,\ldots,>_n\rangle$ in $\mathbb{N}^2$ and outputs 
$\langle >_1,\ldots,>_{i-1},>_{i+1},\ldots,>_n\rangle$.

A deficit reduction is a 4-tuple $(P,c,e,d)$ where $P$ is a 
preference profile, $c$ and $d$ are candidates, and $e$ is an edit 
such that $\deficit(P,c,d) > \deficit(P\swap e,c,d)$.
The full sequence of deficit reductions with respect to candidate $c$
 over a sequence of edits 
$\langle e_1,\ldots, e_p\rangle$ on a preference profile $P$, denoted 
$D(P,c,\langle e_1,\ldots,e_p\rangle)$, is the 
nonrepeating sequence of deficit reductions 
$\langle (P_1,e_{i_1},c,d_1),\ldots,(P_q,e_{i_q},c,d_q)
\rangle$ of maximum length
such that, for all $k \in \{1,\ldots,q\}$, $P_k = P\swap e_{i_i} \swap
 \cdots \swap e_{i_k -1}$, $\deficit(P_k,c,d_k) > 
\deficit(P_k\swap e_{i_k},c, d_k)$, and for all $j \in \{1,\ldots,k-1\}$, 
$i_j \leq i_k$.

We now define, using the terms given above, Dodgson and Young's scoring
rules.
Let $\mathcal{S}$ be the collection of all sequences of swaps. The 
Dodgson score of candidate $c$ in profile $P$ is the smallest $p \in \mathbb{N}$ such that 
\[(\exists \langle e_1,\ldots, e_p\rangle \in \mathcal{S})[\deficit(P\swap e_1\swap \cdots\swap e_p,c) = 0].\]
Let $\mathcal{D}$ be the collection of all sequences of deletions. 
The Young score of candidate $c$ in profile $P$ is the smallest $p \in 
\mathbb{N}$ such that 
\[(\exists \langle e_1,\ldots, e_p\rangle \in \mathcal{D})[\deficit(P\swap e_1\swap \cdots\swap e_p,c) = 0].\]

\subsection{The generic framework }
Below is a generic algorithm for the voting rules we study and approximate.
Here, $\mathcal{E}$ is a collection of ``legal'' sequences of edits, 
whose exact makeup depends on 
the particular scoring rule in question.  The variable $\mathcal{E}$
is implemented as a priority queue, where priority is given
to sequences of edits $S'$ that, when applied to the preference
profile $P$, have the fewest edits per deficit reduction,
i.e., that minimize $|S'|/|D(P,c,S')|$. We call this
quantity the marginal cost of $S'$. We
define $|S'|/|D(P,c,S')| = \infty$ whenever $|D(P,c,S')| = 0$.

$S$ is a list of edits made.

In order to emphasize the key components of this algorithm, we have omitted
important but mundane steps. For instance, the algorithm needs to
compute $\deficit(P,c)$. We will discuss such details when we discuss
the actual Dodgson---and Young---approximation rules.

\vspace{.5cm}

\parbox{8cm}{
\texttt{
\noindent
Input: A preference profile $P$ and a candidate $c$.
  \begin{enumerate}
    \item let $S = \langle \rangle$
  \item while $\deficit(P,c) > 0$
   \item \label{enum:pull} \tab let $S' = \argmin_{S'' \in \mathcal{E}} |S''|/|D(P,c,S'')|$ 
     \item \tab let $\langle e_1, \ldots, e_p\rangle = S'$
     \item \tab  let $P = P \swap e_1\swap \cdots\swap e_p$
       \item \tab concatenate$(S,\langle e_1, \ldots, e_p\rangle)$
\item \label{enum:end} output $|S|$
  \end{enumerate}}}
\section{Approximating Dodgson's scoring rule}
For any candidate $c$, we say that a sequence of swaps $s_1,\ldots,s_p$ is 
\emph{$c$-normal} on $P$
if, for each $k \in \{1,\ldots, p\}$, $c$ is involved in swap
$s_k = (i,j)$ on $P\swap s_1 \swap\cdots\swap s_{k-1}$ and $c(<_i) = j-1$.

Let $P$ be a preference
profile and let $\mathcal{E}'$ be 
the collection of all $c$-normal swap sequences where, for each sequence, 
there is a single voter's preference list to which all swaps in the 
sequence apply. Note then that every such sequence has a distinct
last element, so we can represent each sequence in $\mathcal{E}'$ by
storing its last element only. Let us call the voting rule based on the
generic algorithm with $\mathcal{E} = \mathcal{E}'$
``Marginal-Cost-Greedy-Dodgson.''
\begin{theorem}\label{t:dodg-time}
  The running time of Marginal-Cost-Greedy-Dodgson, when $\mathcal{E} = \mathcal{E}'$,
is $\bigO(N^2\log N)$, where $N$ is the length of the input.
\end{theorem}
\begin{proof}
  Let $(P,c)$ be the input to the algorithm, where $\mathcal{E} = 
\mathcal{E}'$ and $P$ has $m$ candidates and $n$ voters. We first need 
to initialize the data structures
used. It takes linear time to calculate $\deficit(P,c,d)$ on all $d \in 
C\setminus\{d\}$ (note that we can compute $\deficit(P,c)$ at the same
time).  
Next we need to initialize $\mathcal{E}'$. There are at most $n(m-1)$ 
sequences $S'$ in $\mathcal{E}'$, and there are at most 
$m(m-1)/2$ distinct values for $|S'|/|D(S')|$ that any such sequence can take.
 So (regarding $\mathcal{E}$ as a priority queue)
it takes $\bigO(\log m)$ comparisons to add any such sequence (which
we recall is represented by the last element of the sequence) to
$\mathcal{E}$. Note that we can calculate $|S'|/|D(S')|$ for every
sequence $S'$ in $\mathcal{E}$ in a single pass through $P$. The worst case
is when $n$ is as small as possible, so the worst case running time
for initialization is $\bigO(N\log N)$

After initialization, the algorithm performs swaps on $P$ until $c$
is the Condorcet winner. Note that any given swap is performed at
most once. For each swap applied, the algorithm must remove
the corresponding swap from the queue (since whenever a swap is applied
it follows that
the swap sequence ending with that swap has also been applied),
and it must update the marginal cost of each swap sequence remaining in 
$\mathcal{E}$ that applies to the current voter's preferences. 
Thus, every swap may require $\bigO(m)$ updates to $\mathcal{E}$. 
Assuming that all swaps in $\mathcal{E}$ sharing a common voter
are connected via a linked list, each update can happen in constant
time. As during initialization, the worst case for these procedures
occurs when $n$ is as small as possible, so the running time for 
this part of the algorithm is 
$\bigO(N^2)$ 

 Finally, every time a swap causes the deficit against some
opponent to go from positive to zero the entire queue needs to be
reprioritized, which means we must pass through all swap sequences and
recalculate  This can happen at most $(m-1)$ times. Again, the worst-case 
running time is when $n$ is as small as possible, so it is $\bigO(N^2\log N)$.

\end{proof}

We turn now to the approximation bound. Our proof assumes 
there is a $c$-normal Condorcet sequence of swaps witnessing the
Dodgson score of $c$. The following proposition shows that our
assumption is valid.
\begin{proposition}
 For every preference profile $P$ and candidate $c$ there is a $c$-normal 
Condorcet swap sequence of length equal to the Dodgson score of $c$. 
\end{proposition}
\begin{proof}
Let $p$ be the Dodgson score of $c$ and $\langle s_1,\ldots, s_p\rangle$ be a 
Condorcet swap sequence with respect to candidate $c$
on preference profile $P = \langle >_1,\ldots >_n\rangle$. Let 
$\langle >'_1,\ldots, >'_n\rangle = P\swap s_1\swap\cdots\swap s_p$. 
Choose $i \in V$ and let $\langle s_1',\ldots, s_q'\rangle$
be the subsequence of $\langle s_1,\ldots, s_p\rangle$ consisting of all 
swaps on voter $i$'s preferences. Let $d' = 
\argmax_{d \in C : c >'_i d}(\w{d}{>_i} - \w{c}{>_i})$. Since 
it requires at least $\w{d'}{>_i} - \w{c}{>_i}$ swaps in order for 
$c >'_i d'$ to hold, it must be the case that $|\langle s_1',\ldots, 
s_q'\rangle| \geq \w{d'}{>_i} - \w{c}{>_i}$. 
So, removing from $S$ each swap in $\langle s_1',\ldots, s_q'\rangle$ 
and appending the sequence $\langle (i,\w{c}{>_i}+1),\ldots, (i,\w{d'}{>_i})\rangle$ yields a Condorcet sequence that has no more swaps than $S$ originally had.
\end{proof}

\begin{theorem}\label{t:dodg-approx}
  Marginal-Cost-Greedy-Dodgson is an $(\ln m + 1)$-approximation of Dodgson score, where 
$m$ is the number of candidates in the input election.
\end{theorem}
\begin{proof}
Let $P$ be a preference profile over $m$ candidates and $n$ voters and
let  $c$ be a candidate in $\{1,\ldots,m\}$. Let $x$ be the Dodgson
score of $c$ on $P$ and let $S^*$ be a $c$-normal Condorcet
sequence of $P$. Let $y = \deficit(P,c)$ and let 
$\langle (P^*_1, c, s^*_1, d^*_1),\ldots, (P^*_y,c,s^*_y,
d^*_y)\rangle = \dfr{c}{P}{S^*}$. Let $S$ be
the same as in the algorithm on input $(P,c)$ at the time 
line~\ref{enum:end} is reached (i.e., it is the sequence of all
swaps the algorithm applies to $P$),
and let $\langle (P_1, c, s_1, d_1),\ldots, (P_y,c,s_y,d_y)
\rangle = \dfr{c}{P}{S}$. 

The basic idea behind our proof is that the number of deficit
reductions in a sequence that witnesses the Dodgson score of 
$c$, such as $S^*$, is equal to the number of deficit reductions
in the sequence $S$ that the algorithm produces. So to compare $|S|$ to $|S^*|$ 
we partition the swaps in $S$ (respectively, $S^*$) among the
deficit reductions and then match the deficit reductions
in $S$ to those in $S^*$.
The partitioning is easy: For $S$ it is just the marginal cost associated 
with each deficit
reduction. For $S^*$ we fudge the marginal cost in a straightforward
way. The matching and the order in which matched elements are compared
are the trickiest parts of the proof.

For every $k \in \{1,\ldots,y\}$, let $r(s_k)$ denote the marginal
cost the algorithm associates with $s_k$ (i.e., $|S'|/|D(P,c,S')|$,
where $S'$ and $P$ are as in line~\ref{enum:pull} during the iteration
when the algorithm chooses $s_k$ to be in $S'$). Clearly,
\[|S| = \sum_{k=1}^y r(s_k).\]

Let $\sigma$ denote a permutation over 
$\{1,\ldots,y\}$ that satisfies the 
following constraints. 
\begin{enumerate}
\item  For every $j \in \{1,\ldots,y\}$, $d^*_j =  d_{\sigma(j)}$.
\item For every $j,k \in \{1,\ldots,y\}$,
 if $s^*_{k} = s_{j}$ then $k = \sigma(j)$.
\end{enumerate}
Clearly, such a mapping exists.

 For each $i \in \{1,\ldots,n\}$, let $S^*_i$ (respectively, $D^*_i$) 
be the subsequence of all swaps in $S^*$ (respectively, $\langle s^*_1, 
\ldots, s^*_y\rangle$) that apply to voter $i$ only (i.e., all swaps that
for some $j$ are of the form $(i,j)$). Let $p = |D^*_i|$ and let 
$D_i=\langle s_{k_1},\ldots,s_{k_p}\rangle$ be the subsequence of all swaps 
in $\langle  s_1,\ldots, s_y\rangle$ that $\sigma$ maps to
some element in $D^*_i$. In particular, this subsequence preserves the
order in which the algorithm applies the swaps.

 We claim, for every $q\in\{1,\ldots,p\}$, that  $r(s_{k_q}) \leq 
|S^*_i|/(|D^*_i|+1-q)$. This is because, by our
construction of $\sigma$,
 at the time the algorithm is about to choose $s_{k_q}$ it has not 
chosen $s^*_{\sigma(k_q)}$ nor any of the other swaps in $S^*_i$ that come 
after it (in fact, the algorithm may not have chosen a single swap in $S^*_i$). 
Because 
the subsequence $\langle s_{k_1},\ldots,s_{k_p}\rangle$
preserves the order in which the swaps were made, the algorithm still 
needs at this point to close deficits against the 
candidates $d_{k_q},d_{k_q+1}\ldots,d_{k_p}$ 
$(= d^*_{\sigma(k_q)},d^*_{\sigma(k_q+1)},\ldots,d^*_{\sigma(k_p)})$. 

So at the time the algorithm chooses swap $s_{k_q}$, it could instead take
the longest subsequence of $S^*_i$ that remains unchosen. Obviously,
this subsequence is at most $|S^*_i|$ swaps long and, as discussed above,
it yields at least $|D^*_i|+1-q$ deficit reductions. Since $s_{k_q}$ was
chosen as part of a sequence $S'$ for which $|S'|/|D(P,S',c)|$ ($=r(s_{k_q})$, 
where $P$ here is taken to be in the same state as when $S'$ was chosen) was as small
as possible, our claim holds. But then 
\begin{eqnarray*}
|S| &=&\sum_{k=i}^y r(s_k)\\
 &\leq& \sum_{i=1}^n\sum_{q=1}^{|D^*_i|}|S^*_i|/(|D^*_i|+1-q)\\
&\leq&\sum_{i=1}^n\sum_{q=1}^m|S^*_i|/(m+1-q)\\
&\leq&|S^*| \ln m + 1
\end{eqnarray*}

\end{proof}
\section{Approximating Young's scoring rule}
For a given preference profile $P$, let $\mathcal{E}''$ be the 
collection of all single-element sequences of deletions on $P$.
Let us call the voting rule based on the generic
algorithm with $\mathcal{E} = \mathcal{E}''$
``Marginal-Cost-Greedy-Young.''
\begin{theorem}
  Marginal-Cost-Greedy-Young runs in time $\bigO(N^2\log N)$.
\end{theorem}
\begin{theorem}
   Marginal-Cost-Greedy-Young is a $\bigO(\log m)$ approximation of
the Young score, where $m$ is the number of candidates in a given 
input preference profile.
\end{theorem}
The proofs of the above theorems are essentially analogous to those
of theorems~\ref{t:dodg-time} and~\ref{t:dodg-approx}.
\section{Conclusion}
We provide scoring rules that approximate Dodgson's
and Young's rules to within logarithmic factors. 
Assuming $\P \neq \NP$, the bound on Dodgson's scoring rule is
within a constant factor of the
optimal polynomial-time approximation. 
Many natural questions arise from this work. What are the
actual optimal polynomial-time approximations to Dodgson
and Young scores, assuming $\P \neq \NP$? How frequently
do the final candidate rankings according to our scoring rules 
equal those given by Dodgson's and Young's rules on the
same input? 

Our paper gives a general framework for edit-based scoring
rules, in which different types of edits could be combined
to produce an endless stream of distinct voting rules. The 
basic problem of comparing, in such a broadened setting,
edit distances against the edit sequences
produced by the kind algorithms presented here
seems worthy of further research.

Finally, in the introduction we explained our voting rules in 
terms of an auction-like mechanism, where we assumed that all voters value all
edits equally. This suggests an intriguing line of study: What if
that is not how voters feel? For instance, it seems
natural to us that voters would be less willing to make swaps
higher up on their preference lists, and so would require a 
higher price to make them. And in many settings we would
expect the value placed on edits to vary across a 
population of voters. So how
would allowing voters to specify the cost of each edit affect
the score our algorithm produces, compared to the 
corresponding edit-distance-based score?
\bibliographystyle{alpha}
{

 \bibliography{dodgbib,jbib}

}
\end{document}